\documentclass[twocolumn,superscriptaddress,preprintnumbers,amsmath,amssymb]{revtex4}
\usepackage{amsmath}
\usepackage{amssymb}
\usepackage{dcolumn}
\usepackage{graphicx}
\usepackage{bm}
\usepackage{epstopdf}
\usepackage{threeparttable}
\usepackage{float}
\usepackage{color}
\usepackage{mathtools} 

\begin{document}
\title{On the Concept of Static Structure Factor}
\author{Kai Zhang} 
\email{kai.zhang.statmech@gmail.com}
\affiliation{Department of Chemical Engineering, Columbia University, New York, New York, 10027, USA}

\date{\today}

\begin{abstract}
We clarify the confusion in the expression of the static structure factor $S({\bf k})$ in the study of condensed matters and discuss its explicit form  that can be directly used in calculations and computer simulations.

\end{abstract}

 \maketitle



For a $N$-particle system of volume $V$ with density profile $\rho({ \bf r})= \sum_{i=1}^N \delta({\bf r} - {\bf r_i})$, the static structure factor $S({\bf k})$ is defined as the Fourier density correlation~\cite{hansen:2013}
\begin{align}
S({\bf k}) &\equiv \frac{1}{N}\left\langle \hat{\rho}_{\bf k}  \hat{\rho}_{\bf -k} \right\rangle, 
\end{align}
where the Fourier transform pair is $\hat{\rho}_{\bf k}  = \int \rho({\bf r}) e^{i {\bf k} \cdot {\bf r}} d {\bf r} = \sum_{i=1}^N e^{i {\bf k} \cdot {\bf r_i}}$ and $\rho({ \bf r})= (2 \pi)^{-3}\int \hat{\rho}_{\bf k}   e^{-i {\bf k} \cdot {\bf r}} d {\bf k} $~\cite{footnote1}. A general expression of $S({\bf k})$ used in computer simulations is 
\begin{align}
S({\bf k}) &=\frac{1}{N} \left\langle \sum_{i=1}^N e^{i {\bf k} \cdot {\bf r_i}} \sum_{j=1}^N e^{-i {\bf k} \cdot {\bf r_j}}  \right\rangle \nonumber \\
\Aboxed{ &= \frac{1}{N} \left\langle  \left | \sum_{i=1}^N  \cos ({\bf k} \cdot {\bf r_i})  \right|^2 +  \left | \sum_{i=1}^N  \sin ({\bf k} \cdot {\bf r_i})  \right|^2 \right\rangle.}
\end{align}
In Equation (2), the summation of  cosine or sine function scales with $N^2$ (so $S({\bf k}) \sim N$), when ${\bf k}=0$ or the density profile  $\rho({\bf r})$ has a periodicity matching the wavevector {\bf k} (e.g.  a crystalline structure). Therefore, Equation (2) is valid for both amorphous structures (e.g. uniform liquids, glasses) and periodic structures (microphases, crystals). 

For simulations  in a cubic box of finite size $L=\sqrt[3]{V}$ with periodic boundary conditions (PBCs), one sets the wavevector increment to be $\Delta k = 2\pi / L$, because PBCs impose a constraint $L$ on the maximum possible period. Any period larger than $L$ (any wavevector smaller than $\Delta k$) is unphysical. One can then choose to calculate $S(\bf k)$ for  $(n_k+1)^3$ number of ${\bf k}$'s, $\Delta k (n_x, n_y, n_z)$ with $n_x,n_y,n_z=0,1,2,\cdots,n_k$~\cite{allen:1987}. Since it is often desirable to express $S(\bf k)$ as a function of the modulus $k = | \bf k |$, one can group ${\bf k}$'s that have the same magnitude $k$ (or fall in the same bin [$k, k+dk$] in case of  finite numerical accuracy) and use the averaged value for $S(k)$.

In the statistical mechanics theory of liquids,  $S(\bf k)$ is also related to the radial distribution function $g(\bf r)$ or the pair correlation function $h({\bf r})=g({\bf r})-1$~\cite{footnote2} by the Fourier transform. However, among several widely used textbooks, two apparently conflicting expressions are used. For instance, references~\cite{allen:1987,chandler:1987,hansen:2013,gotze:2009} use
\begin{align}
S({\bf k}) &=  1+ \rho \int d{\bf r} g({\bf r})   e^{i {\bf k} \cdot {\bf r}} (\equiv 1+\rho \hat{g}_{\bf k} )
\end{align}
while references~\cite{barrat:2003,mcquarrie:2000,debenedetti:1996,wales:2003} use
\begin{align}
S({\bf k}) &=  1+ \rho \int d{\bf r}  (g({\bf r})-1)   e^{i {\bf k} \cdot {\bf r}} (\equiv 1+\rho \hat{h}_{\bf k} ).
\end{align}
We argue here that, although Equation (3) is conceptually correct,  it is practically not useful and could be misleading. Since $\lim\limits_{r\to\infty} g({\bf r})=1$ is {\em not} absolutely integrable, the integral $\hat{g}_{\bf k}= \int d{\bf r} g({\bf r})   e^{i {\bf k} \cdot {\bf r}}$ in (3) does not converge (or the Fourier transform  $\hat{g}_{\bf k}$ does not exist)~\cite{gotze:2009}. The problem is solved by using $h({\bf r})=g({\bf r})-1$ instead, whose Fourier transform $\hat{h}_{\bf k}= \int d{\bf r} (g({\bf r})-1)   e^{i {\bf k} \cdot {\bf r}}$ is a well-defined converging integral. The original diverging part (or $N$-dependent part in the case of finite box size) in (3) is then extracted in a separate term of delta function~\cite{hansen:2013},
\begin{align}
S({\bf k}) &=  1+ \rho \int d{\bf r}  (g({\bf r})-1)   e^{i {\bf k} \cdot {\bf r}}+ N\delta_{{\bf k},0},
\end{align}
where  the relation $ \int d{\bf r}    e^{i {\bf k} \cdot {\bf r}}=(2\pi)^3 \delta({\bf k})=V \delta_{{\bf k},0}$ is used~\cite{footnote3,footnote4}.
Therefore, Equation (4) is correct and a special case of (5)  when ${\bf k} \neq 0$. If the system is uniform, Equation (4) can be further simplified to~\cite{wales:2003}
\begin{align}
\Aboxed{
S({k}) 
&=  1+ 4\pi \rho \int_0^{\infty} d{r}  (g({r})-1)  r^2\frac{\sin(kr)}{kr}, ~~(k\neq0).}
\end{align}

In principle, we can  obtain $g({\bf r})$ from inverse Fourier transform of $S({\bf k})$. However, Equation (3) will lead to a confusing result $g({\bf r}) = (2\pi)^{-3} \int d{\bf k}  (S({\bf k})-1)/\rho  e^{-i {\bf k} \cdot {\bf r}}$~\cite{allen:1987,hansen:2013}, if the integration interval is not carefully chosen. In fact, because $\hat{g}_{\bf k} = \hat{h}_{\bf k}+ (2\pi)^3 \delta({\bf k})$ has a $\delta$-function singularity  at ${\bf k}=0$, like $S({\bf k}$), the full  integration $(2\pi)^3\int d{\bf k} \hat{g}_{\bf k} e^{-i {\bf k} \cdot {\bf r}}=g({\bf r})$ conceptually~\cite{footnote4,footnote5}, while $(2\pi)^3\int_{{\bf k}\neq0} d{\bf k} \hat{g}_{\bf k} e^{-i {\bf k} \cdot {\bf r}}=g({\bf r})-1$.   Because $\hat{h}_{\bf k}$ only has a {\em finite} discontinuity at ${\bf k}=0$~\cite{footnote4},    inverse Fourier transform of $\hat{h}_{\bf k}$ from Equation (4) converges  and gives the more useful result~\cite{debenedetti:1996,wales:2003}
\begin{align}
g({\bf r}) &=  1+ \frac{1}{(2\pi)^3} \int d{\bf k}  \frac{S({\bf k})-1 }{\rho}  e^{-i {\bf k} \cdot {\bf r}} \\
&=  1+\frac{1}{2\pi^2} \int_0^{\infty}  d{k}  \frac{S({k})-1 }{\rho}  k^2\frac{\sin(kr)}{kr}. ~({\rm uniform})
\end{align}
Note that the above integration  has been extended to include ${\bf k}=0$, in which $\lim\limits_{{\bf k}\to0} S({\bf k}) $, instead of $S(0)$, is used at ${\bf k}=0$~\cite{footnote4}.

\begin{figure}
\begin{center}
\includegraphics[width=0.49\columnwidth]{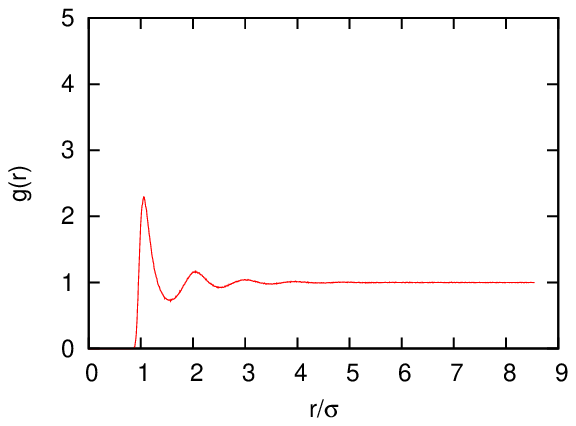}
\includegraphics[width=0.49\columnwidth]{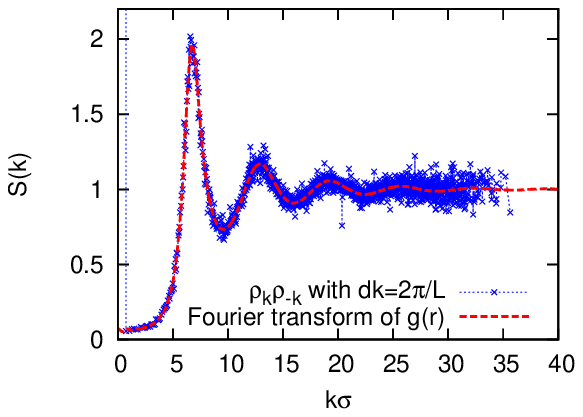}
\includegraphics[width=0.49\columnwidth]{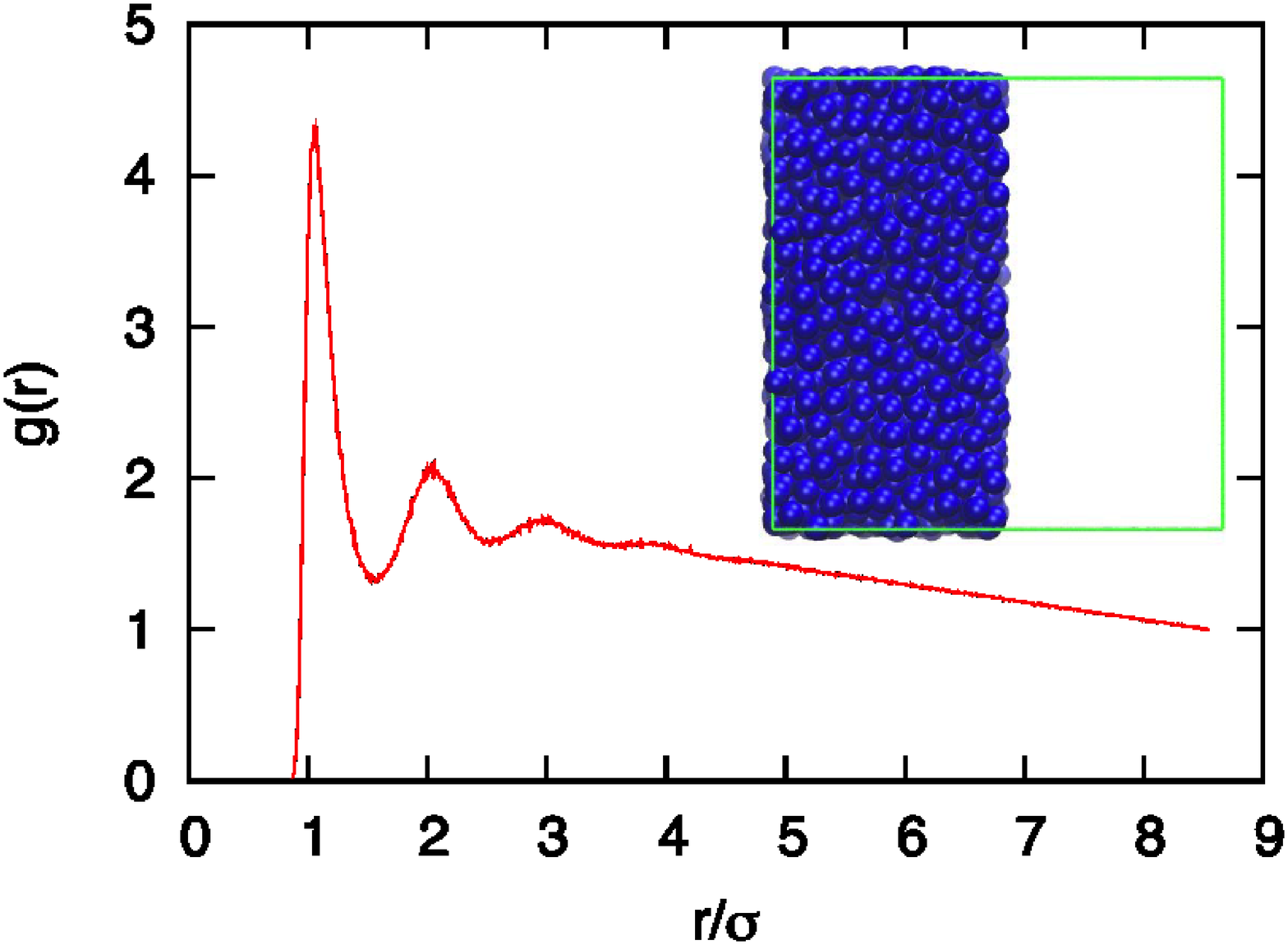}
\includegraphics[width=0.49\columnwidth]{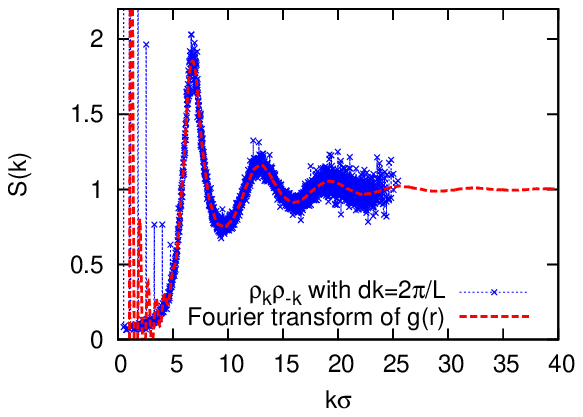}
\includegraphics[width=0.49\columnwidth]{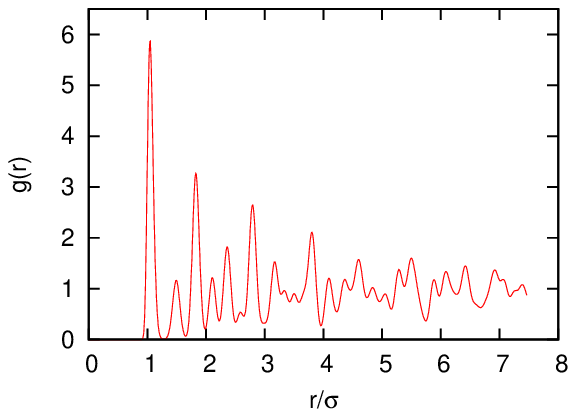}
\includegraphics[width=0.49\columnwidth]{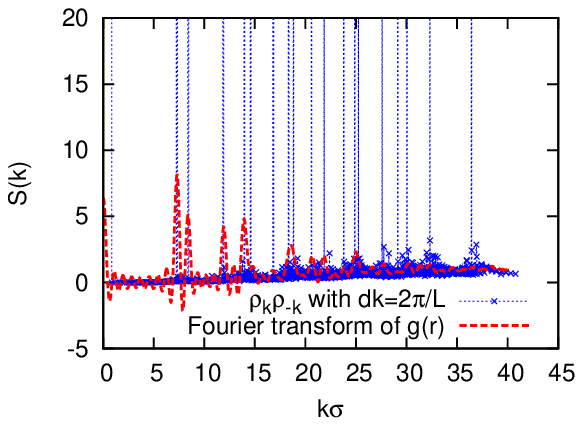}
\caption{ $g(r)$ (left) and $S(k)$ (right) for Lenard-Jones (of core size $\sigma$) liquid at density $\rho\sigma^3=0.8$ (top), lamellar phase at density $\rho\sigma^3=0.4$ (middle)  and face-centered cubic crystal at density $\rho\sigma^3=1.2$. For $S(k)$ both direct calculation using Equation (2) (blue crosses) and Fourier transform of $g(r)$ using Equation (6) (red line) are shown.}
\label{fig:1}
\end{center}
\end{figure}

Finally, we test the direct method in Equation (2) and the Fourier transform method in Equation (6)  by calculating $S(k)$ of Lennard-Jones systems in  liquid, lamellar and crystalline state (Fig.~\ref{fig:1}). For liquid structures (top), the two methods agree with each other very well. For the lamellar phase (middle), amorphous structural features above $k\sigma > 5$ are still preserved. In the small $k$ limit, strong oscillations of $S(k)$ from the Fourier method and high peaks from the direct method are resulted from the long wavelength period of the lamellae.  But the direct method gives more quantitative measurement of the long-range order, including the major peak $S(k^*)=270.55$ ($\sim$$N$) at $k^* = 2\pi/L=0.36744$ for this $N=2000$ system (out of the axis range in Fig.~\ref{fig:1}). For crystalline structures (bottom), the Fourier transform of $g(r)$ only capture {\em some} peak locations of $S(k)$. The actual $S(k)$ of crystals should have peak height that scales with system size $N$ and peak locations that correspond to those in X-ray crystallography (blue vertical lines) . The Fourier transform in Equation (6) should not apply to modulated phases or crystals because it assumes isotropic structures. The large value $S(0)=N$, due to the discontinuity of $S({\bf k})$~\cite{footnote4}, can be obtained from the direct method by setting ${\bf k}=0$ in Equation (2). The physically meaningful $\lim\limits_{k \to 0}S(k)=\rho k_B T \kappa$, where  $\kappa$ is the isothermal compressibility~\cite{hansen:2013,footnote4}.

\begin{acknowledgments}
We thank Till Kranz and Thomas Witelski for helpful discussions.
\end{acknowledgments}


\begin{thebibliography}{11}
\bibitem[{foo({\natexlab{b}})}]{footnote2}
\bibinfo{note}{In some textbooks, the symbol $g(r)$ is used for the pair
  correlation function instead of $h(r)$.}
\end{thebibliography}

\end{document}